\documentclass[jmp,aps,showpacs,reprint]{revtex4-1}
\usepackage{graphicx}
\usepackage{dcolumn}
\usepackage{amsmath}
\usepackage{epsfig}
\RequirePackage{xspace}

\usepackage{relsize}

\newcommand{\BR}{\mathcal{B}}

\begin{document}

\title{
\large \bfseries \boldmath Semileptonic Decay of
$D_s^+\to \pi^0 \ell^+ \nu_\ell$ Via Neutral Meson Mixing }

\author{Hai-Bo Li}
\affiliation{Institute of High Energy Physics, P.O.Box 918,
Beijing  100049, P.R. China }
\author{Mao-Zhi Yang}\thanks{Corresponding author}\email{ yangmz@nankai.edu.cn}
\affiliation{School of Physics, Nankai University, Tianjin 300071,
P.R. China}

\date{\today}


\begin{abstract}
Based on the mechanism of neutral meson mixing, we predict
the branching fraction of the semileptonic decay of $D_s^+\to \pi^0 \ell^+ \nu_\ell$ ($\ell=e$, $\mu$)
using recently measured branching fraction of $D_s^+\to \eta e^+\nu_e$ by
BESIII experiment . We also give a formula that can describe the neutral meson mixing of
$\pi^0$, $\eta$, $\eta'$ and a pseudoscalar gluonium $G$ in a unified way.
The predicted branching fraction of $D_s^+\to\pi^0 e^+\nu_e$ decay is
$\BR(D_s^+\to\pi^0 e^+\nu_e)=(2.65\pm 0.38)\times 10^{-5}$.
 It is important to search for the decay $D_s^+\to \pi^0 \ell^+ \nu_\ell$ in experiment, in order to understand the mechanism of $\pi^0$ production in $D_s^+\to\pi^0$ transition
in semileptonic decay. We also estimate  the branching fraction of
$D_s^+\to \pi^0\tau^+\nu_\tau$, which is the only kinematically allowed semi-tauonic decay mode of the charmed meson,  since the mass value of $D_s^+$ meson is just slightly above
the threshold of $\pi^0\tau^+$ generation in the semi-tauonic decay.

\end{abstract}

\maketitle
In principle neutral mesons with hidden flavors can mix via strong and electromagnetic interactions
if these mesons carry the same quantum numbers, such as spin, parity and charge conjugation
that are exactly conserved in strong and electromagnetic interactions. For vector mesons with
$J^{PC}=1^{--}$, there are $\rho-\omega$ \cite{coon1,mal1,mal2,con,gard,terasaki} and
$\omega-\phi$ mixing \cite{bena1,kucu,bena2,gronau1,gronau2}. For pseudoscalar mesons of
$J^{PC}=0^{-+}$, there are $\pi^0 -\eta$ \cite{coon1,coon2}, $\eta -\eta'$ \cite{feld1,bena1,feld2,ric}
and $\eta'$-gluonium mixing \cite{ric}. Meson mixing is an interesting phenomenon that can be used to
explain some specific decay processes of heavy mesons. For example, both $\omega -\phi$ mixing and weak annihilation are
used as the mechanism that leads to the semileptonic decay of $D_s^+\to \omega  e^+\nu_e$ \cite{gronau2}.
Due to the mixing of $\omega -\phi$ mesons, there is a small component of $s\bar{s}$ in the wave
function of $\omega$ meson, therefore the transition of $D_s^+\to \omega$ can be induced by the $s\bar{s}$ component
in $\omega$ via $D_s^+ \to (s\bar{s}) \ell^+ \nu_\ell$ transition.
Meanwhile the weak annihilation mechanism refers to the effect that one $\omega$ meson is preradiated
nonperturbatively from the $c\bar{s}$ system in strong interaction, then the $c\bar{s}$ system annihilates into $e^+\nu_e$
via the charged weak current. The theoretical analysis of Ref. \cite{gronau2} shows that, if the value
of the branching fraction of $D_s^+\to \omega  e^+\nu_e$ exceeds $2\times 10^{-4}$, the nonperturbative
weak annihilation effect rather than $\omega -\phi$ mixing would be important in the decay.

We consider the semileptonic decay of $D_s^+\to \pi^0\ell^+\nu_\ell$ in this work. Similar to the decay
process of $D_s^+\to \omega  e^+\nu_e$, the process $D_s^+\to \pi^0\ell^+\nu$ can only occur via
$\pi^0 -\eta$ mixing and the nonperturbative weak annihilation effects. For the weak annihilation effect,
it occurs by the preradiation of a $\pi^0$ meson from the $c\bar{s}$ system in $D_s^+$ meson, followed by
the weak transition of $c\bar{s}\to e^+\nu_e$.  However, compared to $D_s^+\to \omega  e^+\nu_e$ decay, the
weak annihilation effect in $D_s^+\to \pi^0  e^+\nu_e$ is doubly suppressed because the nonperturbative
radiation of $\pi^0$ is suppressed by not only the Okubo-Zweig-Iizuka (OZI) rule \cite{OZI1,OZI2,OZI3}, but also isospin violation. So the weak annihilation effect in $D_s^+\to \pi^0  e^+\nu_e$  decay is relatively small. We can neglect the weak annihilation effect and only consider $\pi^0-\eta$ mixing contribution in the following analysis.
We also give the estimation for the order of the branching fraction of $D_s^+\to \pi^0 \tau^+\nu_\tau$, which is 
the only semi-tauonic decay mode of $D_s^+$ meson and is highly suppressed
by its tiny phase space, since the mass value of $D_s^+$ meson is just slightly above
the threshold of $\pi^0\tau^+$ generation in the semileptonic decay.
We hope that these semileptonic decays will be searched for in experiment, so that the decay dynamics can be further investigated.

Now that $\pi^0-\eta$, $\eta -\eta'$ and $\eta'-G $ can mix in each pair, then in principle they can mix
in an enlarged unified way. The mixing between $\pi^0$, $\eta$, $\eta'$ and the gluonium $G$ should be
described uniformly. Let us define $\pi^0_q=\frac{u\bar{u}-d\bar{d}}{\sqrt{2}}$, $\eta_q=\frac{u\bar{u}+d\bar{d}}{\sqrt{2}}$, $\eta_s=s\bar{s}$, and $G$ as the pure pseudoscalar gluonium.
The physical mesons of $\pi^0$, $\eta$, $\eta'$ and a pseudoscalar
$\eta_G$ should be the mixing states of these flavor bases
\begin{equation}
\begin{pmatrix}
\pi^0\\ \eta \\ \eta' \\ \eta_G
\end{pmatrix}
=V
\begin{pmatrix}
\pi^0_q\\ \eta_q \\ \eta_s \\ G
\end{pmatrix}\label{eq1},
\end{equation}
where $V$ is a $4\times 4$ unitary matrix that describes the mixing between the pseudoscalar bases
$\pi^0_q$, $\eta_q$, $\eta_s$ and $G$. The matrix $V$ is treated as a real matrix here for simplicity.
For the unitarity of the mixing matrix $V$,
there should be 6 independent parameters in the $4\times 4$ matrix $V$. Let's set $i=1,\; 2,\; 3,\; 4$ for
$\pi^0_q$, $\eta_q$, $\eta_s$ and $G$, then the 6 free parameters can be denoted as $\theta_{12}$,
$\theta_{23}$, $\theta_{34}$, $\theta_{13}$, $\theta_{14}$ and $\theta_{24}$, which can be viewed as mixing
angles. About the mixing angles, the following statements should be given:

1) The mixing angle $\theta_{12}$
is for the mixing between $\pi^0_q$ and $\eta_q$, which is isospin-violating, therefore the mixing
angle $\theta_{12}$ should be small,  here we  denote it as $\theta_{12}\equiv\delta$;

2) $\theta_{23}$
is the mixing angle for $\eta_q-\eta_s$, which is denoted as $\theta_{23}\equiv\phi$;

3) $\theta_{34}$ is the angle for
$\eta_s-G$ mixing, which is denoted as $\theta_{34}\equiv\phi_G$;

4) $\theta_{13}$ is the angle for the mixing of
$\pi^0_q-\eta_s$. The mixing between these two states is not only isospin-violating, but also
with larger mass-gap between these two states. So the mixing angle should be  tiny,
which can be neglected, $\theta_{13}\sim 0$;

5) $\theta_{14}$ is
the mixing angle for $\pi^0_q-G$ mixing, which is also isospin-violating and with larger mass-gap between these two mixing
states, so the mixing angle is also tiny and can be set to $\theta_{14}\sim 0$;

6) $\theta_{24}$ is the angle  for the mixing of $\eta_q-G$.
We find that, if considering $\theta_{24}$ to be very small, then
the component of $G$ in $\eta$ will be tiny, which is consistent with the result of
QCD sum rule calculation that the coupling of the gluonium to $\eta$ is much smaller than its
coupling to $\eta'$ \cite{def}. So we can take $\theta_{24}\sim 0$ in the following analysis for simplicity.

Then the nonzero mixing angles considered in our scenario are
\begin{equation}
\theta_{12}\equiv\delta,\;\;\; \theta_{23}\equiv\phi,\;\;\; \theta_{34}\equiv\phi_G.
\end{equation}
With these three nonzero mixing angles, we can write the sub-mixing matrices explicitly as
\begin{equation}
V1=\begin{pmatrix}
    \cos\delta & -\sin\delta &0 & 0\\
    \sin\delta & \cos\delta  &0 & 0\\
    0         & 0          &1 & 0\\
    0         & 0          &0 & 1
   \end{pmatrix}
\end{equation}
for $\pi^0-\eta_q$ mixing,
\begin{equation}
V2=\begin{pmatrix}
     1 & 0       &0         & 0\\
     0 & \cos\phi & -\sin\phi & 0\\
     0 & \sin\phi & \cos\phi  & 0\\
     0 & 0       &0         & 1
   \end{pmatrix}
\end{equation}
for $\eta_q-\eta_s$ mixing, and
\begin{equation}
V3=\begin{pmatrix}
     1 & 0 &0          & 0\\
     0 & 1 &0          & 0\\
     0 & 0 &\cos\phi_G  & \sin\phi_G &\\
     0 & 0 &-\sin\phi_G & \cos\phi_G &
    \end{pmatrix}
\end{equation}
for $\eta_s-G$ mixing. The total mixing matrix can then be taken as
\begin{eqnarray}
V &=& V_3 V_1 V_2 \nonumber\\
 &=& \begin{pmatrix}
    \cos\delta & -\sin\delta \cos\phi &  \sin\delta \sin\phi & 0\\
    \sin\delta &  \cos\delta \cos\phi & -\cos\delta \sin\phi & 0\\
    0         &  \cos\phi_G \sin\phi & \cos\phi_G \cos\phi & \sin\phi_G\\
    0         &  -\sin\phi_G \sin\phi & -\sin\phi_G \cos\phi & \cos\phi_G
 \end{pmatrix} \label{v-matrix}
\end{eqnarray}
Considering that $\delta$ should be tiny, so we can make the approximation:
$\cos\delta\sim 1$, $\sin\delta\sim \delta$. Then Eq.(\ref{v-matrix}) can be
simplified as
\begin{equation}
V\simeq\begin{pmatrix}
    1      & -\delta \cos\phi &  \delta \sin\phi & 0\\
    \delta &  \cos\phi        & -\sin\phi        & 0\\
    0      &  \cos\phi_G \sin\phi & \cos\phi_G \cos\phi & \sin\phi_G\\
    0      &  -\sin\phi_G \sin\phi & -\sin\phi_G \cos\phi & \cos\phi_G
 \end{pmatrix} \label{v-matrix2}
\end{equation}
Substituting Eq. (\ref{v-matrix2}) into Eq. (\ref{eq1}), we can obtain
\begin{eqnarray}
&&\mid \pi^0\rangle =\mid \pi^0_q\rangle-\delta \cos\phi\mid \eta_q\rangle +\delta \sin\phi\mid \eta_s\rangle,
      \label{pi-w}\\
&&\mid \eta\rangle =\delta\mid \pi^0_q\rangle+ \cos\phi\mid \eta_q\rangle - \sin\phi\mid \eta_s\rangle, \label{eta-w}
\end{eqnarray}
\begin{eqnarray}
\mid \eta'\rangle &=& \cos\phi_G \sin\phi\mid \eta_q\rangle +\cos\phi_G \cos\phi\mid \eta_s\rangle\nonumber\\
 &&+\sin\phi_G\mid G\rangle, \label{etap-w}\\
\mid \eta_G\rangle &=& -\sin\phi_G \sin\phi\mid \eta_q\rangle -\sin\phi_G \cos\phi\mid \eta_s\rangle\nonumber\\
 &&+\cos\phi_G\mid G\rangle. \label{fG-w}
\end{eqnarray}
Eq. (\ref{pi-w}) implies that the physical neutral pion $\pi^0$ is dominantly $\pi^0_q$ which is a component with
isospin 1, and there are small components of $\eta_q$ and $\eta_s$ mixed in $\pi^0$. Note $\delta$ is a tiny quantity.
Eq. (\ref{eta-w}) indicates that $\eta$ meson is mainly $\eta_q$ and $\eta_s$ with a small component of $\pi^0_q$ in
it, and the gluonium component can be neglected. Eq. (\ref{etap-w}) shows that $\eta'$ is a mixing state of $\eta_q$,
$\eta_s$ and the gluonium component $G$. This is consistent with the expression given in Ref. \cite{ric}. Finally
the state $\eta_G$ in Eq. (\ref{fG-w}) is the orthogonal state of $\eta'$, which is also mixing state of $\eta_q$,
$\eta_s$ and the gluonium component $G$. If the mixing angle $\phi_G$ is small, then $\eta_G$ is dominantly a gluonium
state.

The mixing angle $\delta$ can be determined by the ratio of the branching fractions of $\eta'\to\pi^+\pi^-\pi^0$ and
$\eta'\to\pi^+\pi^-\eta$ decays, where the former is $G$-parity violating, which
can only occur via $\pi^0-\eta$ mixing \cite{gro}. The dominant decay mode of $\eta'$ is $\eta'\to\pi^+\pi^-\eta$.
With the mixing scheme given in Eqs. (\ref{pi-w}) and (\ref{eta-w}), the decay amplitudes of $\eta'\to\pi^+\pi^-\pi^0$
and $\eta'\to\pi^+\pi^-\eta$ are
\begin{eqnarray}
&&\;\;\;\;\langle \pi^+\pi^-\pi^0\mid H \mid \eta'\rangle \nonumber\\
&&= \langle\pi^+\pi^-\pi^0_q\mid H \mid \eta'\rangle
         -\delta\cos\phi\langle\pi^+\pi^-\eta_q\mid H \mid \eta'\rangle \nonumber\\
&& \;\;+\delta\sin\phi\langle\pi^+\pi^-\eta_s\mid H \mid \eta'\rangle ,   \label{etap-pi}
\end{eqnarray}
\begin{eqnarray}
&&\;\;\;\;\langle \pi^+\pi^-\eta\mid H \mid \eta'\rangle \nonumber\\
&&= \delta\langle\pi^+\pi^-\pi^0_q\mid H \mid \eta'\rangle
        +\cos\phi\langle\pi^+\pi^-\eta_q\mid H \mid \eta'\rangle \nonumber\\
&& \;\;-\sin\phi\langle\pi^+\pi^-\eta_s\mid H \mid \eta'\rangle ,\label{etap-eta}
\end{eqnarray}
where $H$ is the Hamiltonian that induces the $\eta'$ three-body decays. The matrix element
$\langle \pi^+\pi^-\pi^0_q\mid H \mid \eta'\rangle$ is $G$-parity violating, so we can take
$\langle \pi^+\pi^-\pi^0_q\mid H \mid \eta'\rangle\sim 0$. Then Eqs. (\ref{etap-pi}) and (\ref{etap-eta})
indicate
\begin{equation}
\frac{\langle \pi^+\pi^-\pi^0\mid H \mid \eta'\rangle}{\langle \pi^+\pi^-\eta\mid H \mid \eta'\rangle }=-\delta.
\end{equation}
So we can obtain the ratio of the decay branching fractions
\begin{eqnarray}
&&\;\;\;\;\frac{\BR(\eta'\to \pi^+\pi^-\pi^0)}{\BR(\eta'\to\pi^+\pi^-\eta) }\nonumber\\
&&=\left| \frac{\langle \pi^+\pi^-\pi^0\mid H \mid \eta'\rangle}
{\langle \pi^+\pi^-\eta\mid H \mid \eta'\rangle}\right| ^2
\frac{\phi_s(\eta'\to \pi^+\pi^-\pi^0)}{\phi_s(\eta'\to\pi^+\pi^-\eta) }\nonumber\\
&&=\delta^2 \frac{\phi_s(\eta'\to \pi^+\pi^-\pi^0)}{\phi_s(\eta'\to\pi^+\pi^-\eta)} \label{delta-br},
\end{eqnarray}
where $\phi_s(\eta'\to \pi^+\pi^-\pi^0(\eta))$ is the phase space volume of the decay mode
$\eta'\to \pi^+\pi^-\pi^0(\eta)$. The ratio of the phase space can be calculated directly to be  \cite{gro,cheng}
\begin{equation}
\frac{\phi_s(\eta'\to \pi^+\pi^-\pi^0)}{\phi_s(\eta'\to\pi^+\pi^-\eta) }=17.0. \label{phase-v}
\end{equation}
The branching fraction of $\eta'\to \pi^+\pi^-\pi^0$ has been measured by CLEO collaboration in 2008 \cite{cleo} and
BESIII collaboration in 2012 and 2017 \cite{BESIII1, BESIII2}.
The relative ratio of $\BR(\eta'\to \pi^+\pi^-\pi^0)/\BR(\eta'\to\pi^+\pi^-\eta)$ was analyzed based on the recent data
of BESIII, and its value is determined to be $(8.8\pm 1.2)\times 10^{-3}$ as  in Ref. \cite{fang}. With this ratio from experiment, and Eqs. (\ref{delta-br}) and (\ref{phase-v}), we can obtain
\begin{equation}
\delta^2=(5.18\pm 0.71)\times 10^{-4}. \label{delta-value}
\end{equation}

Next we shall go on to discuss the semileptonic decays of $D_s^+\to \pi^0 \ell^+\nu_\ell$ and $D_s^+\to \eta \ell^+\nu_\ell$.
According to Eqs. (\ref{pi-w}) and (\ref{eta-w}), both of these decays occur via the $\eta_s$ component in
$\pi^0$ and $\eta$ at tree level. For semileptonic decay of $D_s^+$ to a pseudoscalar $P$, the hadronic matrix element
involved in the decay amplitude is
\begin{eqnarray}
&&\;\;\;\; \langle P(p_2)\mid V_\mu\mid D_s(p_1)\rangle\nonumber\\
&&=F_+^{D_sP}(q^2)(p_1+p_2-\frac{m^2_{D_s}-m_P^2}{q^2})_\mu\nonumber\\
&&\;\;\;\;+F_0^{D_sP}(q^2)\frac{m^2_{D_s}-m_P^2}{q^2}q_\mu, \label{matrix-el}
\end{eqnarray}
where $q=p_1-p_2$, and $p_{1,2}$ are the momenta of $D_s^+$ and $P$ mesons, respectively.
$m_{D_s}$ and $m_P$ are the masses of $D_s^+$ and $P$, respectively.
$F_{+}^{D_sP}(q^2)$ is the so-called
vector form factor, and $F_{0}^{D_sP}(q^2)$ the scalar form factor. To avoid the divergence in the hadronic
matrix element in Eq. (\ref{matrix-el}) as $q^2\to 0$, there should be
\begin{equation}
 F_{+}^{D_sP}(0)=F_{0}^{D_sP}(0).
\end{equation}
The differential decay width of a semileptonic decay $D_s^+\to Pl^+\nu$ can be calculated to be
\begin{eqnarray}
\frac{d\Gamma}{dq^2}(D_s^+&\to&P \ell^+\nu_\ell)=\frac{G_F^2|V_{cs}|^2}{24\pi^3m^2_{D_s}q^4}(q^2-m_\ell^2)^2|\vec{p}_P|\nonumber\\
&& \times\left[ (1+\frac{m_\ell^2}{2q^2})m^2_{D_s}|\vec{p}_P|^2|F_+^{D_sP}(q^2)|^2\right. \nonumber\\
&&+\left.\frac{3m_\ell^2}{8q^2}(m^2_{D_s}-m^2_P)^2|F_0^{D_sP}(q^2)|^2\right], \label{width-ml}
\end{eqnarray}
where $\vec{p}_P$ is the momentum of the pseudoscalar $P$ in the rest frame of $D_s^+$ meson. $G_F$ is the Fermi
constant, $m_\ell$ the lepton mass, and $V_{cs}$ the Cabibbo-Kobayashi-Maskawa (CKM) matrix element. For $\ell=e$ and $\mu$,
$m_\ell^2$ can be neglected. Then
\begin{equation}
\frac{d\Gamma}{dq^2}(D_s^+\to Pl^+\nu)=\frac{G_F^2}{24\pi^3}|V_{cs}|^2|F_+^{D_sP}(q^2)|^2|\vec{p}_P|^3. \label{width-e}
\end{equation}
For the $q^2$-dependent form factor $F_+^{D_sP}(q^2)$, we consider the modified pole model \cite{bech}
\begin{equation}
F_+^{D_sP}(q^2)=\frac{F_+^{D_sP}(0)}{(1-\frac{q^2}{m^2_{\rm pole}})(1-\alpha\frac{q^2}{m^2_{\rm pole}})}, \label{ffq2}
\end{equation}
where $\alpha$ is a free parameter, and $m_{pole}$ is fixed to be the mass of the vector state $D_s^{*+}$.
The BESIII collaboration has measured the branching fraction of $D_s^+\to \eta^{(\prime)}e^+\nu_e$ and the $q^2$-dependent form factor $F_+^{D_s\eta^{(\prime)}}(q^2)$ \cite{BESIII3}. The parameter $\alpha$ was fitted
for each decay mode. The value of $\alpha$ for $F_+^{D_s\eta}(q^2)$ is \cite{BESIII3}
\begin{equation}
\alpha=0.304(44)(22).
\end{equation}

According to Eqs. (\ref{pi-w}) and (\ref{eta-w}), both the transition matrix elements of
$\langle \pi^0\mid V_\mu\mid D_s^+\rangle$ and $\langle \eta\mid V_\mu\mid D_s^+\rangle$
can be related to $\langle \eta_s\mid V_\mu\mid D_s^+\rangle$ by the following relation
\begin{eqnarray}
\langle \pi^0\mid V_\mu\mid D_s^+\rangle &=& \delta \sin\phi \langle \eta_s\mid V_\mu\mid D_s^+\rangle , \\
\langle \eta\mid V_\mu\mid D_s^+\rangle &=& -\sin\phi \langle \eta_s\mid V_\mu\mid D_s^+\rangle,
\end{eqnarray}
because both the transitions of $D_s^+\to \pi^0$ and $D_s^+\to \eta$ occur through the component $s\bar{s}$
mixed in $\pi^0$ and $\eta$ mesons. Then we can get the relations between the form factors
\begin{eqnarray}
F_{+,0}^{D_s\pi}(q^2)&=&\delta\sin\phi F_{+,0}^{D_s\eta_s}(q^2),\\
F_{+,0}^{D_s\eta}(q^2)&=&-\sin\phi F_{+,0}^{D_s\eta_s}(q^2).
\end{eqnarray}
Comparing both sides of the above two equations, we have the following relation
\begin{equation}
\frac{F_{+,0}^{D_s\pi}(q^2)}{F_{+,0}^{D_s\eta}(q^2)}=-\delta. \label{ff-ratio}
\end{equation}
Note that the form factors only explicitly depend on $q^2$, not on the masses of the initial and final
mesons.

Using the expression of the differential decay width in Eq. (\ref{width-e}) and considering the $q^2$-dependent form factor in Eq. (\ref{ffq2}), we can get the ratio
\begin{eqnarray}
&&\;\;\;\;\frac{\BR(D_s^+\to\pi^0 e^+\nu_e)}{\BR(D_s^+\to\eta e^+\nu_e)}=\delta^2\times\nonumber\\
&&\frac{\int_0^{(m_{D_s}-m_{\pi})^2}dq^2 |\vec{p}_\pi|^3/
      [(1-\frac{q^2}{m^2_{D_s^{*+}}})(1-\alpha\frac{q^2}{m^2_{D_s^{*+}}})]^2}
      {\int_0^{(m_{D_s}-m_{\eta})^2}dq^2 |\vec{p}_\eta|^3/
      [(1-\frac{q^2}{m^2_{D_s^{*+}}})(1-\alpha\frac{q^2}{m^2_{D_s^{*+}}})]^2},\nonumber\\
&& \label{br-ratio}
\end{eqnarray}
where $F_{+,0}^{D_s\pi}(0)/F_{+,0}^{D_s\eta}(0)=-\delta$ is used according to Eq. (\ref{ff-ratio}). Using Eqs.  (\ref{br-ratio}), (\ref{delta-value}) and the measured branching fraction by BESIII \cite{BESIII3}
$\BR(D_s^+\to\eta e^+\nu_e)=(2.323\pm 0.063\pm 0,063)\%$, we can obtain the branching fraction of $D_s^+\to\pi^0 e^+\nu_e$
decay
\begin{equation}
\BR(D_s^+\to\pi^0 e^+\nu_e)=(2.65\pm 0.38)\times 10^{-5}, \label{br-prediction}
\end{equation}
where the error mainly comes from the uncertainties of the parameters $\alpha$, $\delta^2$, and the error of the
experimentally measured branching fraction of $D_s^+\to\eta e^+\nu_e$. The error caused by the uncertainty of the parameter $\alpha$ is about 1.3\%, while the error caused by the uncertainty of $\delta^2$ is about 13.6\%, and the uncertainty caused by the error of the experimental value of $\BR( D_s^+\to\eta e^+\nu_e)$ is about 3.8\%.  The error caused by the other sources is tiny
which can be ignored.

The prediction in Eq. (\ref{br-prediction}) is based on the contribution of $\pi^0 -\eta$ mixing scheme given in
Eqs. (\ref{pi-w}) and (\ref{eta-w}), and the possible weak annihilation contribution is neglected.
As analyzed before, the weak annihilation contribution is doubly suppressed because it
both violates isospin invariance and suppressed by the OZI rule. Therefore the weak annihilation contribution
must be small. So the prediction or at least
the order given in our prediction in Eq. (\ref{br-prediction}) is reliable.  Therefore, measurement of the branching fraction
of $D_s^+\to\pi^0 e^+\nu_e$ in experiment  can be used to test any sizable contribution from the weak annihilation effect.

Next we shall go on to consider the decay  of $D_s^+\to\pi^0\tau^+\nu_\tau$, in which the mass of $D_s^+$ is just
slightly above the threshold of $ \pi^0 \tau^+$  production. It is suppressed by both the small mixing amplitude
of $\pi^0 -\eta$ and the limited phase space in the decay. In addition, the decay of $D_s^+\to\pi^0\tau^+\nu_\tau$ is the only kinematically allowed decay for the charmed mesons.  Therefore, it is interesting to know the order of
the decay rate of the process $D_s^+\to\pi^0\tau^+\nu_\tau$.

According to Eq. (\ref{width-ml}), the decay width of $D_s^+\to P\tau^+\nu_\tau$ involves not only the form factor
$F_+^{D_s P}(q^2)$, but also the form factor $F_0^{D_s P}(q^2)$. The form factor $F_0^{D_s P}(q^2)$ can not be measured
in experiment through the semileptonic decay process of $D_s^+ \to P \ell^+ \nu_\ell$.  Since there is no any
information on the form factor $F_0^{D_s P}(q^2)$ in experiment up to now, we still use the modified pole model
for the $q^2$-dependence of $F_0^{D_s P}(q^2)$. Specifically, for $P=\pi^0$, the form factor $F_0^{D_s \pi^0}(q^2)$ is
taken as
\begin{equation}
F_0^{D_s\pi}(q^2)=\frac{F_0^{D_s\pi}(0)}{(1-\frac{q^2}{m^2_{\rm pole}})(1-\beta\frac{q^2}{m^2_{\rm pole}})},
\label{ff0q2}
\end{equation}
where $m_{\rm pole}$ should be a state of $c\bar{s}$ system with $J^P=0^+$, which can be taken as the mass
of $D_{s0}^*(2317)^+$, $m_{D_{s0}^*(2317)}=2317.7\pm 0.6$ MeV according to PDG \cite{PDG}, and $\beta$ is a free
parameter.

\begin{center}
\begin{figure}[h]
\epsfig{file=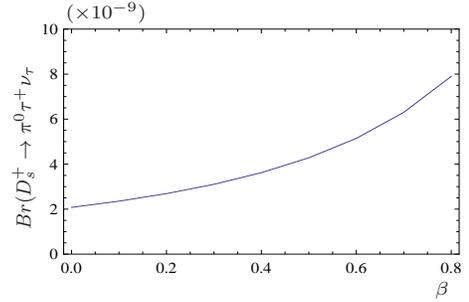,width=6cm,height=4cm} \caption{The branching fraction of $D_s^+\to\pi^0\tau^+\nu_\tau$
varying with the parameter $\beta$, where the other parameter $\alpha$ and the branching fraction of
$D_s^+\to\eta e^+\nu_e$ are input from the centeral values of experimental measured results.} \label{fig1}
\end{figure}
\end{center}

Using Eqs. (\ref{width-ml}) and (\ref{width-e}), we can obtain the ratio of the branching fractions of the decay modes of
$D_s^+\to\pi^0\tau^+\nu_\tau$ and $D_s^+\to\eta e^+\nu_e$
\begin{equation}
\frac{Br(D_s^+\to\pi^0\tau^+\nu_\tau)}{Br(D_s^+\to\eta e^+\nu_e)}=\delta^2\frac{a}{b}, \label{br-ratio-tau}
\end{equation}
with
\begin{eqnarray}
a&=&\int_{m_\tau^2}^{(m_{D_s}-m_\tau)^2}dq^2 \frac{(q^2-m_\tau^2)^2}{m^2_{D_s}q^4}|\vec{p}_\pi|\nonumber\\
&& \times\left[ (1+\frac{m_\tau^2}{2q^2})\frac{m^2_{D_s}|\vec{p}_P|^2}
{[(1-\frac{q^2}{m^2_{D_s^{*+}}})(1-\alpha\frac{q^2}{m^2_{D_s^{*+}}})]^2}\right. \nonumber\\
&&+\left.\frac{3m_\tau^2}{8q^2}\frac{(m^2_{D_s}-m^2_\pi)^2}
{[(1-\frac{q^2}{m^2_{D_{s0}^*(2317)}})(1-\beta\frac{q^2}{m^2_{D_{s0}^*(2317)}})]^2}\right]
\end{eqnarray}
and
\begin{equation}
b=\int_0^{(m_{D_s}-m_{\eta})^2}dq^2 \frac{|\vec{p}_\eta|^3}{
      [(1-\frac{q^2}{m^2_{D_s^{*+}}})(1-\alpha\frac{q^2}{m^2_{D_s^{*+}}})]^2}
\end{equation}
Here in deriving Eq. (\ref{br-ratio-tau}), $F_+^{D_s\pi}(0)=F_0^{D_s\pi}(0)$ and $F_+^{D_s\pi}(0)/F_+^{D_s\eta}(0)=-\delta$
have been used. Since there is no any information on $\beta$ in experiment yet, $\beta$ is treated as a free parameter
in this work. As an illustration, the branching fraction of $D_s^+\to\pi^0\tau^+\nu_\tau$ varying with the parameter $\beta$
is depicted in Fig. \ref{fig1} according to Eq. (\ref{br-ratio-tau}), where all the other parameters are input with their central values.

It can be known from Fig.\ref{fig1} that the order of the branching fraction of $D_s^+\to\pi^0\tau^+\nu_\tau$ is
about $10^{-9}$. If one assumes that the parameter $\beta$ is approximately the same order as $\alpha$, say
$\beta$ can be 0.2$\sim$0.4, then the branching fraction of $D_s^+\to\pi^0\tau^+\nu_\tau$ will be
$(2.7\sim 3.6)\times 10^{-9}$.

According to the future physics program at BESIII,  about 6 fb$^{-1}$ integrated luminosity will be collected at the
center-of-mass energy of 4180 MeV at BEPCII \cite{li2017, Ablikim:2019hff}.  For our predicted branching fraction of
$\BR(D_s^+\to\pi^0 e^+\nu_e)=(2.65\pm 0.38)\times 10^{-5}$,  about a few signal events are expected to be reconstructed at BESIII experiment based on the double-tag technique~\cite{Ablikim:2018jun}. We also hope that significant signal will be observed at Belle-II experiment~\cite{Kou:2018nap} and the future super-tau-charm factory~\cite{Luo:2018njj} and , which will collect about 100 times the amount of the current data set at BESIII.
As for the decay of $D_s^+\to\pi^0 \tau^+\nu_\tau$, the decay rate is predicted to be $(2.7\sim 3.6)\times 10^{-9}$,  and is  not yet experimentally observable.

In summary, we study the mixing of $\pi^0-\eta-\eta'-\rm{Gluonium}$ and the mixing scheme is given
in a unified way. Then the branching fractions of $D_s^+\to\pi^0 e^+\nu_e$ is predicted to be
$\BR(D_s^+\to\pi^0 e^+\nu_e)=(2.65\pm 0.38)\times 10^{-5}$, which can be searched for at the BESIII experiment and will be important observable at the future super-tau-charm factory.
It will be interesting to search for the decay $D_s^+\to\pi^0 e^+\nu_e$, in order to understand the decay dynamics, namely to validate the $\pi^0-\eta$ mixing effect and the weak annihilation contribution.
We also estimate the order of the branching fraction of $D_s^+\to\pi^0 \tau^+\nu_\tau$ decay,
which is about $(2.7\sim 3.6)\times 10^{-9}$.  This is the only allowed semi-tauonic decay mode in the charm sector.

\vspace{0.5cm}
This work is supported in part by the National Natural Science Foundation of China under
Contracts No. 11875168, 11375088, 11935018, 11875054;  the Chinese Academy of Sciences under Contract No. QYZDJ-SSW-SLH003.


\end{document}